# A model for individual quantal nano-skyrmions

J.P.Gauyacq[1] and N.Lorente[2,3]

[1] Institut des Sciences Moléculaires d'Orsay (ISMO), CNRS, Univ. Paris-Sud, Université Paris-Saclay, Bât. 520, F-91405 Orsay, France
[2] Centro de Fisica de Materiales CFM/MPC (CSIC-UPV/EHU),
Paseo Manuel de Lardizabal 5, 20018 Donostia-San Sebastian, Spain
[3] Donostia International Physics Center (DIPC),
Paseo Manuel de Lardizabal 4, 20018 Donostia-San Sebastian, Spain

## Abstract

A quantal model description of a discrete localized skyrmion singularity embedded in a ferromagnetic environment is proposed. It allows discussing the importance of various parameters in the appearance of a quantal skyrmion singularity. Analysis of the skyrmion reveals a few specific quantal properties: presence of a whole series of skyrmion states, non-classical nature of the local spins, presence of superposition states and presence of extra-skyrmion states due to the quantization of the central spin of the singularity. The interaction of an electron, tunneling or substrate, with the skyrmion is also described allowing a new view at the origin of the skyrmion stability as well as the possibility to discriminate between ferromagnetic and skyrmion phase in an Inelastic Electron Tunneling Spectroscopy (IETS) experiment, based on the skyrmion quantal properties.

# 1. Introduction

Magnetic skyrmions are local stable arrangements giving rise to special spin textures: typically, the central spins of the skyrmion point upward, whereas the peripheral spins point down. They attracted much attention due to their topological protection, following the original theoretical work of Skyrme[1]; indeed, one cannot deform continuously a skyrmion to reach e.g. the ferromagnetic phase. The early observations of magnetic skyrmions[2,3,4,5,6,7] prompted a series of experimental and theoretical works (see e.g. reviews in [8,9]). Magnetic skyrmions have been proposed for future magnetic memories[10,11,12,13] due to the stabilization provided by the topological protection and to their easy motion under an electric current[14,15,16,17]. Their observation at room temperature has also been reported[18,19]. In this context, the experimental observation[20,21] by spin-polarized STM (Scanning Tunneling Microscopy) of individual localized skyrmions as single singularities in a ferromagnetic environment and the demonstration of the possibility of switching at will between a localized pinned skyrmion and the ferromagnetic phase by injecting a tunneling electron into the system is of particular significance. Indeed, that proves the possibility of writing and deleting pinned skyrmions, the first step towards a skyrmion based magnetic memory. Besides this tunneling electron mechanism, the reversible switch between ferromagnetic and skyrmion state has also been discussed via fast magnetic field pulses[22] and electric field action[23].

Discussing small pinned skyrmions brings in several questions. First, the topological aspects are related to a continuous field description of the system. However, small pinned skyrmions are made of a finite number of discrete atomic spins. Then what happens of the topological protection when one considers a very small skyrmion with a very limited number of spins? This question has been addressed in several theoretical works that consider discrete skyrmion structures[24,25,26,27,28]. Second, another question comes from possible quantal effects in the case of a small geometrical size of the skyrmions and/or of low spin values of the constituting atoms. Indeed, most of the earlier theoretical approaches have considered a classical description of the spin assembly in which the local spins have a fixed length and orientate according to their interactions with the surroundings. Some discussions of the skyrmion quantal dynamics have been presented, however considering classical spins (spin vectors of constant length)[29,30,31,32,33], allowing in particular a discussion of the role of quantum tunneling in the skyrmion collapse route[34,31]. Several studies considering quantal structures with finite size S=1/2 lattice have recently appeared[35,36]. Atomic spins at surfaces are associated to small quantum numbers, so quantal effects can be expected from the quantal nature of the local spins. Superposition of different local spin states is possible and this could a priori influence the skyrmion structure and its interaction with tunneling electrons. Finally topology is often invoked to account for the stability of skyrmions; however, the skyrmion spin structures are on top of a solid metal surface and consequently interact with the substrate electron bath. Indeed, a substrate electron colliding with a skyrmion could bring the rupture needed to break the topological protection. Could these collisions, possibly inelastic, limit the skyrmion stability? Until now, the skyrmion-ferromagnetic switch has been discussed using classical theoretical approaches, typically in terms of Arrhenius plots, energy barrier separating the two structures, LLG (Landau-Lifschitz-Gilbert) formalism (see e.g. ref.[26,22]) or quantal tunneling descriptions[31]. To address the above questions, we developed a simple theoretical model describing a quantal nanoskyrmion; it is made of a finite number of quantal atomic spins interacting together. We show below how the quantal skyrmion differs from a classical one by different features: a rich skyrmionic spectrum with extra non-classical states, presence of superposition states and non-rigid length of the atomic magnetic moments. In addition, our model allows to investigate the interaction of a tunneling electron from an STM tip or of a substrate electron with the skyrmion and thus to examine how it is possible to delete/write a skyrmion and how the substrate electrons limit a skyrmion lifetime. Finally, we discuss how an Inelastic Electron Tunneling Spectroscopy (IETS) experiment can reveal the quantal specificities in a nanoskyrmionic system, such as the extra states made possible by the quantal nature of the skyrmion singularity.

## 2. Structure study

The aim of the present work is to design a simple model system, easily tractable computationally, that can be used to discuss the properties of a quantal nano-skyrmion, such as the individual pinned skyrmions revealed recently[20,21]. Various possibilities for the pinning mechanisms have been discussed[37]. Our model describes a nano-skyrmion embedded in a ferromagnetic plane; its atomic structure is schematized in Figure 1. The system is invariant by a π/3 rotation around the z-axis, taken perpendicular to the plane of the structure ($C_6$ symmetry). The seven central spins (one central spin surrounded by a crown of first neighbors spins) are explicitly treated, while the surrounding 24 spins are blocked in the spin-down state along the z-axis and provide a ferromagnetic embedding for the central spins. In the model, we do not introduce any interaction term that would result in skyrmion pinning; the geometry of the system with its ferromagnetic embedding results in the skyrmion localization in the central structure. The nano-skyrmion is pinned at the center of the structure and is confined within the central seven sites, bringing in a very strong size limitation. In the present study, we consider S=3/2 spins. Some tests calculations have been performed for S=1 and S=2 spins, yielding the same qualitative picture.

The Hamiltonian for the model system is written as:

$$H = -J \sum_{i,j} \vec{S_i} \vec{S_j} + AD \sum_i S_{iz}^2 - \sum_{i,j} \vec{D_{ij}} (\vec{S_i} \times \vec{S_j}) + \sum_i B_z S_{iz} - \sum_{i,j} BQ (\vec{S_i} \cdot \vec{S_j})^2$$
$$- K \sum_{i,j,k,l} \{(\vec{S_i} \cdot \vec{S_j})(\vec{S_k} \cdot \vec{S_l}) + (\vec{S_i} \cdot \vec{S_l})(\vec{S_j} \cdot \vec{S_k}) - (\vec{S_i} \cdot \vec{S_k})(\vec{S_j} \cdot \vec{S_l})\}$$

(1)

It is directly inspired from the work of Heinze et al[6,38,39]. $\vec{S_i}$ and $S_{i,z}$ are the local spin operators and their projection on the z-axis for the site i. $J$ is the ferromagnetic exchange interaction only coupling first neighbors in the structure. $AD$ is the magnetic anisotropy of the easy axis type ($AD < 0$). $D_{ij}$ is the Dzyaloshinskii-Moriya (DM) interaction[40,41] between two first neighbors; the $D_{ij}$ vector is taken in the structure plane and perpendicular to the vector joining sites i and j[42,6]. $B_z$ is a small magnetic field along the z-axis introduced to make easier the state assignment. $BQ$ is a bi-quadratic interaction between neighbors and $K$ is the 4-spin interaction coupling the neighboring spins in a diamond (lowest order correction to the Heisenberg model[43]). All interactions between the 7 central spins and the ferromagnetic embedding are included in the Hamiltonian (1).

The Heisenberg interaction, $J$, is taken equal to unity and acts as an energy unit in the model, when describing the role of the various interactions. The magnetic anisotropy term, $AD$, is taken strong enough (equal to 1.5 $J$) to make it difficult to induce a flip of local spins by a single electron.

As discussed in Ref.[6,38], the DM and 4-spin interactions are the key ingredients for the appearance of skyrmionic states. We diagonalized the Hamiltonian (1) in the complete basis set formed of direct products of the eigenstates of $S_{iz}$, for various values of the DM and K interactions. As an example, Figure 2 shows the energies of the lowest states in the system as a function of the DM interaction for a given moderate K interaction (K = 0.1 J). It appears that at small DM, the ground state is ferromagnetic (FM) with all spins aligned down along the z-axis (FM state: red circles). As DM is increased, the FM state crosses a series of states and the ground state becomes a skyrmionic state (Sk state, see below why we label it as a skyrmionic state). Most of the crossings in Fig.2 are real crossings between states of different symmetries with respect to a π/3 rotation around the z-axis (note that the projection of the total spin of the system on the z-axis is not a good quantum number, due to the presence of the DM interaction). The crossing around DM ≈ 1.36 J involves the FM state and a Sk state of a different symmetry and it thus corresponds to a ferromagnetic-skyrmion phase transition of the Landau first order type[44].

Figure 3(a) shows the configuration of local spins in the Sk state that is the ground state at large DM in Fig.2. The interaction parameters are K/J = 0.1 and DM/J = 1.4, i.e. close to the phase transition. The figure presents a cut of the structure in the (y,z) plane; the atoms are located on the y-axis. At each site, we plot a vector whose components are the $<S_{i,u}>$ (u = y, z), mean values for the Sk state of the projections of the $\vec{S_i}$ on the three coordinate axis (for the sites presented in the figure, the mean value of the local spin is in the (y,z) plane). The spins of the FM embedding are presented in green and the spins treated explicitly in red. The central spin is almost (99 %) in a pure $m_z$ = +3/2 state, pointing opposite to the spins in the FM embedding. The intermediate spins in the crown around the center span the transition between the singularity at the center and the surrounding FM phase. This spin structure is exactly the one that is observed in pinned nano-skyrmions by spin-polarized STM[20,21] and thus confirms the appearance in our model of a pinned nano-skyrmion of the Néel Q=1 type. Interestingly, it appears that this picture of the quantal Sk state does not fit that of a classical skyrmion. Indeed, the crown spin vectors in Fig.2 are not vectors of length 3/2 rotated from upward or downward positions, but they suffer a significant length reduction due to quantal state mixing. Their length, $|\langle\vec{S_i}\rangle|$, is around 0.84 in Fig.3a. Each spin in the crown is not an eigenstate of the spin projection on a given direction but a superposition of several eigenstates of the spin projection. The present quantal nano-skyrmion thus appears qualitatively different from the classical skyrmions usually discussed, where one goes from the central singularity to the FM surroundings by a succession of simple spin rotations. As further differences from a classical system, one can see that Fig. 2 exhibits a whole series of Sk states with energies quasi-parallel to that of the lowest Sk state (blue squares in Fig.2). They also correspond to a central singularity with an almost up spin and differ by the state mixings in the crown; their existence is linked to the $C_6$ symmetry of the system, i.e. to the possibility to have a central singularity with arrangements of the spins in the crown of different symmetries with respect to a rotation of π/3 around z-axis.

In addition, the quantal nature of the central spin allows for the existence of a singularity of skyrmionic type with a significant weight on spin configurations with a central spin which is neither in the up nor down state. Violet triangles in Fig.2 show states with a large weight on configurations with a central $m_z$ = +1/2 state. Indeed, such states are impossible with a classical description and are pure quantal objects. Figure 3(b) presents the configuration of local spins in one of these extra quantal skyrmion states. It is the state #21 in figure 2 for the DM = 1.4 J, K = 0.1 J case shown in Fig.3a. As discussed below, this state is of particular significance for the excitation from the ground skyrmion state by a tunneling electron. The central spin is opposite to the FM embedding with a 69% probability to be in the $m_z$ = +1/2 state and 30% in the $m_z$ = +3/2 state. Again for these states, the modulus of the mean value of the local spins appears much smaller than 3/2: $|\langle \vec{S_i} \rangle|$ amounts to 0.8 and 1.0 for the central and crown spins, respectively, in Fig.3b.

Skyrmions are associated with winding properties of the spin texture of the system. Recently, Lohani et al[36] have proposed to examine the W parameter introduced in[45] and related to the winding properties of the structure and defined for a discrete spin lattice by:

$$W = \sum_\Delta \frac{1}{2\pi} \tan^{-1} \left[ \frac{\frac{8}{27} \vec{S_i} \cdot (\vec{S_j} \times \vec{S_k})}{1 + \frac{4}{9}((\vec{S_i} \cdot \vec{S_j}) + (\vec{S_i} \cdot \vec{S_k}) + (\vec{S_k} \cdot \vec{S_j}))} \right]$$

where the summation extends over the triangles mapping the structure. The numerical factors have been adapted to accommodate the S=3/2 case. As shown in Ref.[45], W is the total signed area of the surface defined by the $\vec{S_i}$ vectors. In a classical description with $\vec{S_i}$ vectors of constant length, it is the topological charge of the spin lattice as shown in [45]. As suggested in Ref.[36], we computed the W parameter for the various states found in our system by replacing the classical $\vec{S_i}$ vectors by the mean values of the $\vec{S_i}$ operators for the state under consideration. First, as expected, this parameter vanishes for the ferromagnetic state. We computed the W parameter for the skyrmionic state depicted in Fig.3a; it amounts to 0.386. Similarly for the four lowest skyrmionic states in Fig.2 (for the K/J=0.1 and DM/J=1.4 parameters) the W parameter is between 0.2 and 0.4. These are not integers as discussed in Ref.[36] however this confirms the different structure of the FM and Sk states.

One can go a step further and analyze the *angular structure* of the spin lattice by normalizing the $<\vec{S}_i>$ vectors to 3/2 while retaining their direction. One can thus analyze the winding properties of the system restricted to its angular part. As expected, the W parameter becomes an integer. It amounts to one for all the states labelled skyrmionic above, i.e. we recover the topological charge of a classical skyrmion. Thus the states labelled Sk have the same qualitative spin structure as observed experimentally and if the local spins are normalized they exhibit a topological charge of 1. In this sense, the present Sk states appear in our present model study as the quantal correspondent of the classical skyrmion, the main difference appears to be the varying length of the various $<\vec{S}_i>$ vectors, which is a signature of significant quantal effects.

We also computed the W parameter for the extra skyrmionic states with a central spin dominantly in the $S_z=+1/2$ state. For the state in Fig. 3b it is equal to 0.065 and if the $<\vec{S}_i>$ vectors are normalized, it is equal to 1. This confirms the skyrmionic character of this state, the small value of the W parameter being attributed to the small length of the mean values of the $\vec{S}_i$ vectors in this state (see Fig.3b).

Excitations from the FM state can also be found with energies quasi-parallel to that of the FM state in Fig.2 (green diamonds); there are seven excitations of the magnon type (the projection of the total spin of the system on the z-axis differs roughly by one unit from the FM state); they can be seen as spin waves confined in the finite size structure and adapted to the symmetry of the system. We also computed the W parameters for the magnonic states in Fig2. For all these states, it is not an integer and it is very small in the $10^{-3}$-$10^{-7}$ range, the smallest values being found for the states with the strongest magnonic character (total spin differing by almost one unit from that of the FM state). When the $<\vec{S}_i>$ vectors are normalized, W vanishes. The magnonic states thus have no winding character and, as expected, have a spin texture similar to that of the FM state. The W parameter thus appears as a useful criterion to separate the states in the system in two categories: the FM state and its excitations and the skyrmionic states.

The identification of the various states in Fig.2 is mainly obtained by looking at the dominant configurations of local spins in the state. The lowest lying states in the spectrum are easily identified and associated with well-defined configurations of local spins; however, their characteristics fade out when climbing up in the spectrum due to increasing mixing between the various states (effect of the entanglement of the various spins in the system).

From energy calculations such as those in Fig.2, one can obtain the phase diagram of the system shown in Figure 4. It displays the frontier between regions in the (DM, K) plane where the FM state (below the frontier line) or the Sk state (above the frontier line) is the ground state of the system. The phase diagram confirms that sizeable DM and/or 4-spin interactions are required to stabilize a skyrmion. The diagram is symmetric in DM: indeed a change of sign of the DM interaction only switches between skyrmion and antiskyrmion states, i.e. in the way the spins in the first crown are intermediate between up and down states. In contrast, only the K > 0 interaction favors the skyrmionic phase. One can also notice that in the present model, rather large DM or K interactions are required to reach the phase transition. Their order of magnitude is typically a factor 2.5 above those in Heinze et al[6] study. However, the ab initio study by Meyer et al[46] on Fe/Ir layers on Rh(001) reported rather significant DM and K interactions compared to the Heisenberg exchange. The large couplings required in our model are tentatively attributed to the smallness of the geometrical size of the skyrmion singularity in our model. The small singularity spatial extent requires a rather sharp twist of the spins between the central singularity and the surrounding FM phase, and thus strong skyrmion-favoring interactions. The relation between DM strength and skyrmion size has been discussed in several papers within classical descriptions[47,48] revealing the importance of edge effects (confinement, lattice versus individual skyrmions). We performed some tests calculations with S=1/2 systems with larger numbers of local spins in the central structure. Typically, increasing the number of spins in the central structure from 7 to 16 resulted in a decrease by a factor around 1.5 of the DM interaction required to reach the FM-Sk phase transition, thus confirming the strong role of the system size in accounting for the large interactions needed in the present study.

At small DM interaction, when increasing the 4-spin interaction, the ground state switches from the FM state to a magnon-like state. Skyrmions only appear above a minimal DM interaction, leading to a skyrmion-magnon transition.

The anisotropy interaction (AD term in equation (1)) has also been mentioned as influencing the occurrence of a skyrmion structure. Indeed, a strong easy axis magnetic anisotropy tends to make the FM phase more rigid. Tests calculations revealed that decreasing the magnetic anisotropy favors the skyrmion over the FM phase, however, the effect remains limited as can be seen in Figure 5 which presents the phase diagram of the system (disappearance line of the ferromagnetic phase) for different magnetic anisotropy strengths.

## 3 Excitation of the structure by collision of an electron

An electron (either a tunneling electron from an STM tip or a substrate electron) colliding with the spin structure can induce excitations or de-excitation of the system. We describe these transitions using the strong coupling approach described in [49,50].

We describe the system in the basis set formed by the direct product of the local spin states of the individual atoms:

$$| B_j \rangle = | m_1, m_2, \dots, m_7 \rangle \qquad (2)$$

The local spin states at site i are eigenstates of $S_{i,z}$, the projection on the z–axis of the spin of the atom at site i, $m_i$ is the associated quantum number. Diagonalization of the Hamiltonian (1) in this basis yields the set of magnetic states, $n$, of the system, of energies $E_n$, labelled $| \varphi_n \rangle$ expressed in the $B_j$ basis.

Let us consider the case where the electron is injected in site k of the structure. Electron tunnelling is very fast so that one can neglect the action of Hamiltonian (1) during tunnelling. The incident electron spin then couples to the spin 3/2 of atom k to form a compound state of total spin $S_T$ = 1 or 2. Below we assume that the scattering is dominated by a single symmetry that we chose arbitrarily as $S_T$ = 1 (choosing $S_T$ = 2 would not alter qualitatively the present results). The electron scattering amplitude between tip and substrate then writes as:

$$T_{Tip \to Sub} = \sum_{M_T} | S_T = 1, M_T \rangle T^1_{Tip \to Sub} \langle S_T = 1, M_T | \qquad (3)$$

with the three intermediate states corresponding to the three sublevels, $M_T$, of $S_T$ =1.

Initial (final) states of the collision are direct products of the electron spin state ($\sigma_i$ and $\sigma_f$) by the initial (final) state of the system (states $\varphi_i$ and $\varphi_f$) so that the tunnelling probability for individual states is written as:

$$P(initial \to final)$$

$$= \left| \langle \varphi_f \sigma_f | \left( \sum_{M_T} | S_T = 1, M_T \rangle T^1_{Tip \to Sub} \langle S_T = 1, M_T | \right) | \varphi_i \sigma_i \rangle \right|^2$$

$$= |T^1_{Tip \to Sub}|^2 \left| \sum_{M_T} A^*_{f,M_T} A_{i,M_T} \right|^2$$

i.e. the product of a global tunnelling probability by a series of structure factors describing the coupling/decoupling of the spins of the tunnelling electron and of the system. From these, one can obtain the normalized probabilities for an electron tunnelling between tip and substrate; they only depend on the spin coupling/decoupling coefficient. These probabilities can be used to compute the conductance of the junction if we further assume the global tunnelling probability to be constant over the small energy range scanned by a magnetic IETS experiment.

From this one can obtain the conductance of the system as function of the STM junction bias, $V$, for an unpolarised tip:

$$\frac{dI}{dV} = C_0 \frac{\sum_f Y(eV-E_f) \sum_{\sigma_i,\sigma_f} \left|\sum_{M_T} A_{i,M_T} A^*_{f,M_T}\right|^2}{\sum_f \sum_{\sigma_i,\sigma_f} \left|\sum_{M_T} A_{i,M_T} A^*_{f,M_T}\right|^2} \qquad (4)$$

where $Y(x)$ is the step function and $C_0$ a global conductance. Thus, the relative conductance can be obtained from the only knowledge of the $A_{k,M}$ coupling coefficients.

### 4. Stability of the skyrmion and ferromagnetic structures

The results presented above come from a Hamiltonian diagonalization, so that all states are eigenstates and as such are stable within a description using Hamiltonian (1). The skyrmion state is then stable when it is the ground stable but also when it is an excited state. However, the system described by equation (1) is not alone in space; the spin structure in Fig.1 is adsorbed on a metal surface and the inelastic collision of an electron from the substrate on one of the structure sites can induce a transition from the upper state to the lower state (electron-hole pair creation) and is responsible for the finite lifetime of the excited state (see a discussion of the decay rate induced by electron-hole pair creation in[49,51] and the application to spin decay in[52,53,54,55]). This process can involve thermally excited electrons but the decay is already present at vanishing temperature. The electron-hole pair creation process is discussed below together with its consequences on the skyrmion stability. We concentrate on cases where the system is close to the FM-Sk phase transition, i.e. to situations close to the experimental study of FM-Sk switch[20, 21].

Quantum collapse of a magnetic skyrmion has been discussed in terms of quantum tunneling (QT) and magnon creation[31]. In the present approach dealing with eigenstates, QT would appear as two excited states of neighboring energies and of very different character (a Sk state and a magnon state for example) that are slightly mixed by Hamiltonian (1). Forming a pure Sk state will lead to a slow oscillation between Sk and magnon states that possibly will be transformed into decay by decoherence (see a discussion in [56]). In the present case, this situation cannot occur close to the phase transition, since the FM and lowest Sk states are of different $C_6$ symmetries and thus uncoupled. It can only occur when the lowest Sk state is high lying in the spectrum and close to a magnon state of the same symmetry, i.e. for conditions deep in the FM zone in Fig.4. However, as discussed below, highly excited states in the system are very efficiently quenched by electron-hole pair creation.

Using the method outlined in the previous section, we computed the transition probability between the FM and the lowest Sk state induced by a single collision of an electron on one of the structure sites. In the present case, the FM and Sk states are very different, the local spin states in the various sites are different in the two states and one cannot switch from one state to the other by a simple spin-flip at one site. Furthermore, the flip of the central spin from the up to the down position ($m_z$ = 3/2 to $m_z$ = -3/2) that is required in a FM-Sk transition cannot be induced by a single electron collision due to spin angular momentum conservation and the easy axis magnetic anisotropy is further enhancing this effect[57]. Transition between FM and Sk states is only possible via correlation. The various spins in the system are entangled, so that by touching one site in the structure, all the spins in the system can be modified (see similar situations in[54,58,59,60]). In the present system, correlation between FM-type and Sk-type states exists but is weak and so is the probability for an electron to induce FM-Sk transitions by collision on one of the sites in the crown. It is even vanishing for an electron collision at the center due to symmetry.

Figure 6 presents the direct transition probability, $P_{direct}$(FM↔Sk), induced by a single collision of an electron on one of the crown sites, as a function of the DM interaction, the 4-spin interaction being chosen so that the system is close to the phase transition (the transition probability for an electron injected on the central atom vanishes due to the $C_6$ symmetry). The transition probability is always very weak and decreases when the system is dominated by the DM or by the 4-spins interaction, meaning that the synergic action of DM and 4-spins interactions increases the mixing between states of different characters, i.e. correlation effects.

As discussed in[49,53], due to electron-hole pair excitation, the metastable state has a finite lifetime, $\tau$, that is the inverse of its decay rate, $\Gamma$, and is given by:

$$\frac{1}{\tau} = \Gamma = T(E_F)\frac{\delta E}{h} P_{Direct}(FM \leftrightarrow S_k) \qquad (5)$$

where $T(E_F)$ is related to the substrate electron density at the surface and $\delta E$ is the energy defect between the two states. Using typical value for $T(E_F)$[53] and an energy defect of 1 meV yields lifetime in the 40µs – 40ms range for probabilities in the $10^{-7}$ - $10^{-10}$ range. $\tau$ is inversely proportional to the energy defect that varies with the choice of parameters; however, this confirms the stability of the metastable phase.

We also computed the lifetime of the higher lying excited states. These states are very short lived with lifetimes typically in the $10^{-11}$ s range, except for a few very low lying states. Indeed, in the low energy part of the spectrum, if a state is not the lowest one of the FM type (or of the Sk type) then a one electron interaction can efficiently bring it down to a lower state of the same character, making it short lived (the energy defect of the decay also plays a role, enhancing the lifetime of the lowest states).

Injection of a *low energy* electron by an STM tip located above one of the crown sites will be associated with the probability depicted in Fig.6 and will also be very poorly efficient in inducing a switch between FM and Sk states. However, when the STM bias is increased, other processes involving intermediate excited states come into play: the incident electron brings the system into a high lying excited state, i, that later decays towards lower lying states via electron-hole pair creation. This decay can lead the system back to the ground state but also to the metastable state provided that the intermediate state i has to a certain extent a mixed FM and Sk character (see a similar situation in[58] for transition between Néel states). As mentioned above, when going up in the spectrum, states have a less marked character, thus favoring their role as an intermediate in the indirect Sk-FM transition. Figure 7 shows the indirect probability $P_{ind}$ (Sk→FM) for the DM/J = 1.4 and K/J = 0.1 parameters as a function of the incident electron energy (this is a large DM situation close to the Sk-FM phase transition, however, similar results are obtained along the phase transition line). For these parameters, the Sk state is the ground state and the FM is the second excited. $P_{ind}$ (Sk→FM) sums the contributions from all the Sk → i → FM processes and thus represents the probability for a high energy electron injected from an STM tip to induce a Sk → FM transition. It appears that the indirect process is rather weak at low energy where the excited states have a well-marked Sk or FM character. It becomes more and more efficient as the incident electron energy increases, due to the mixed character of high lying intermediate states and to the increase of the number of intermediate states contributing significantly to the indirect process. One can also notice that the indirect probability depends on the site where the electron is injected, center or crown. At low energy, injecting an electron into the crown is more efficient, possibly due to the high symmetry of the central site, but the effect reverses at larger energy when all states are much mixed. This indirect process then allows quenching the metastable phase using an STM with a large enough bias; the reverse process is also possible and an electron from an STM tip can also create the metastable phase. This kind of process has been observed on pinned skyrmions by Romming et al[20,21].

## 5. Excitation of the system by a tunneling electron

As discussed above, the direct transition between ferromagnetic and skyrmion states induced by collision of an electron is very weak. However, electron coming from an STM tip located above one of the atoms can excite the system to various higher lying states with a significant efficiency, if the junction bias is large enough. As discussed above, the system can be in two different states, ground or metastable. The FM and Sk states are very different and not surprisingly, a tunneling electron can excite both of them but to very different parts of the excited state spectrum, leading to different shapes of the inelastic conductance as a function of bias (IETS, Inelastic Electron Tunneling Spectroscopy).

Let us start with the case of a low DM interaction (0.3 J) and a large 4-spin interaction (0.215 J), such that the system is close to the skyrmion/ferromagnetic phase transition. The ground state is ferromagnetic and the first excited state is a skyrmion state. In the case of systems where $M_z$ (projection of the total spin of the system on the z-symmetry axis) is a good quantum number, the excitation process by collision of an electron follows the $|\Delta M_z| = 0, \pm 1$ selection rule. However, DM interaction mixes states with different $M_z$, ; in the present case of a moderate DM interaction, this mixing is small and the $|\Delta M_z| = 0, \pm 1$ selection rule is active, although in an approximate way.

Figure 8 presents the relative conductance between an STM tip and the system in four cases: for an electron injected into the central atom or into one of the crown atoms, and for the system initially in the ferromagnetic or skyrmion state. The conductance is calibrated to one for large biases above the excitation thresholds. The conductance exhibits a series of steps associated with the opening of excitation processes, and the more efficient is the excitation process, the higher is the step. For site 2 (atom in the crown) and the system initially in the FM state, we observe four well-marked steps, actually associated with seven excitation thresholds with excitation probabilities of a few %. They correspond to magnon states, i.e. states with a $M_z$ differing by around one unit from that of the ferromagnetic state. There are only seven states of that type, built from the FM ground state by a $m_z$ (projection on the symmetry axis of the angular momentum of one of the atoms) change of one unit on one of the atoms of the structure. On site 2, all seven states are excited significantly. In contrast, for site 1 (center atom), the situation is different due to the $C_6$ symmetry around the z-axis which limits the excitation process to only two of the magnon states as seen in Figure 8. The magnon states are quantized in the present finite size system. The situation would be different in a real infinite system where one would observe spin wave excitation.

More interesting is the case of the initial skyrmion state in Figure 8. At low bias, only one high step is present in the conductance for the central atom, i.e. only one excited state has a significant excitation probability. It is one of the extra quantal skyrmion states revealed in the present study, it corresponds to a singularity on the central spin, which is almost in the $m_z$ = +1/2 state (state #16 in the energy spectrum), in contrast to the usual skyrmion which corresponds to the central spin almost in the $m_z$ = +3/2 state. The dominant step in the conductance then corresponds basically to the transition of the central spin from a $m_z$ = +3/2 state to a $m_z$ = +1/2 state, the rest of the system being unchanged. Such a transition is unique and is very efficiently induced by collision with an electron. The conductance for an atom injected into the crown is different: the $m_z$ = 3/2 → $m_z$ = ½ spin flip at the center cannot be induced efficiently by an electron injected into one of the crown atoms. Most of the inelasticity is spread over a large number of states in the 5-15 meV/J range, associated to complex excitations of a skyrmionic state in the crown, i.e. they are associated to excitation in the crown, leaving the central singularity quasi-unchanged.

The shape of the conductance at the center is then characteristic of a skyrmion system, both of the singularity at the center and of the existence of the extra quantal skyrmion states. Changing the size of the skyrmion will not suppress the singularity at the center nor suppress the extra state so that, qualitatively the result obtained in the present finite size model system is expected to survive in larger systems. Indeed the spectroscopic data would change and excitation energies would be different.

Figure 9 presents the analogous of Fig.8 for a larger DM interaction (1.4 J) and a smaller 4-spin interaction (0.1 J). Again, the system is close to the phase transition; the ground state is a skyrmion state as well as the first excited state and the FM state is the second excited state (see Fig.2). The DM interaction being large, the mixing between states of different $M_z$ spin projections is important and the various states of the system appear as important mixing of various spin configurations. This effect increases as one considers higher states in the energy spectrum so that it is difficult to assign a given high lying state to a well-defined spin configuration. This can be linked with the fact that in Figure 2, only the few lowest lying states in each category are labelled, the other having less well-marked characteristics.

Let us consider first in Fig.9 the conductance for the system initially in the ferromagnetic state. Magnon states are low lying and one can recognize a group of steps in the 5-8 meV/J energy range, corresponding to seven partially overlapping thresholds. They are all excited in site 2 and only two of them are excited in site 1, again due to the $C_6$ symmetry axis. They dominate the excitation process for the initial FM state, however not in the overwhelming way seen in the small DM case (Fig.8); excitation to other high lying states are made possible by the mixture of the magnon spin configurations with other spin configurations, and this accounts for the slow rise of the inelastic conductance in the 10-15 meV/J range. So the situation for the excitation of the FM case is similar to that in the low DM case: the dominance of magnon excitation although in a less systematic way due to the strong state mixing.

As for the conductance for the initial state in the ground Sk state (Fig.9), for the central site, all states that are significantly excited are of skyrmion type. One can still recognize excitation to well-marked extra skyrmion states ($m_z = +1/2$ dominant on the central spin) around 8 and 12 meV. However, the mixing induced by strong DM interaction spreads the excitation over several $m_z = +1/2$ dominant states and allows the excitation of other skyrmion states. The conductance for a crown site is very different from that on the central site. On a crown site, all the inelasticity is below 8 meV. All the states that are excited are of the skyrmion type with a very large probability (above 90 %) for the central spin in the $m_z = +3/2$ state. They then correspond to the excitation of the crown spins, the central spin remaining basically in the $m_z = +3/2$ state.

## 6. Discussion and conclusions

The present model of a small central region imbedded in a ferromagnetic environment leads to a phase transition between a ferromagnetic state and a skyrmionic state when the Hamiltonian parameters are changed. The various states are labelled according to the dominant configuration of local spins and to their winding properties using the winding parameter defined in Ref.36 and 45 by replacing the classical spin vectors by the mean values of the spin operators or by normalized mean values. The skyrmionic states in this model appear as the quantal correspondents of the classical skyrmion with the same angular texture but non constant length vectors for the local spins.

The present model of a pinned nano-skyrmion does reproduce some characteristics of those observed experimentally by Romming et al[20,21], in particular the spin arrangement in the structure. For certain values of the DM and 4-spins interactions, the ferromagnetic state is very close in energy from a skyrmionic state and a small parameter change brings in a phase transition. The phase transition can also be induced by varying the magnetic field applied to the system, along the symmetry z-axis (not shown here); indeed, the Sk and FM states do not have the same mean value of the projection of the total spin on the z-axis, so that a variation of B allows to switch the ground state between Sk and FM states.

The switch between FM and Sk can be induced either by a thermal electron from the substrate or by an electron injected by an STM tip. At very low energy, electrons are quite inefficient and as a consequence, at low temperature, the substrate electrons are unable to induce a switch between Sk and FM states and a metastable Sk state (or a metastable FM state) is very long lived. One can stress that this quasi-stability is not linked to a topological protection but to the difficulty of inducing changes in many spins at the same time with a single electron scattering event; only entanglement of the spin sites makes these transitions possible. In the case of an STM experiment at large bias, the possibility of indirect transitions enormously increases the electron efficiency. One can notice, that the transition probabilities obtained in the present model, both for the direct and indirect processes, are significantly larger than those observed experimentally[20,21]. This is tentatively attributed to a size effect; the present model nano-skyrmions involve a small number of spins. Indeed, to switch between Sk and FM states, one needs to flip all the spins in the structure, which is very difficult to perform with a single electron colliding with a single spin. This aspect becomes stronger and stronger when the size (the number of sites involved in the singularity) is increased. The efficiency of the switch depends on the entanglement of the spins in the structure which rapidly decreases as the size increases (independently of decoherence effects[56]). This variation of entanglement (correlation) effects with the system size has already been observed in various spin systems (switch between Néel states[58,59,60], lifetime of spin excitations[54]), so that one can expect significantly smaller transition probabilities for realistic pinned skyrmions, with a number of spins larger than those considered here.

The present model reveals the existence of previously unseen singularities of skyrmionic type that are consequences of the quantal nature of spins and of the symmetry of the spin lattice. Usual classical skyrmions correspond to a central spin polarized in a direction opposite to that of the surrounding spins. A quantal spin S, can be in 2S+1 different states, associated to different projections on the quantization axis and not only in up or down states. We thus found states with a singularity of skyrmionic type, associated with a projection of the central spin on the quantization axis different from +S or –S. These have an energy higher than the usual skyrmionic state and can be excited from the lowest skyrmion; as a consequence, they are not metastable and decay rapidly to the lowest Sk state.

Beyond the possibility to discriminate FM and Sk states with spin-polarized STM imaging[6,20,61], IETS experiments should reveal very different behaviors for the FM and Sk states. Indeed, the FM state leads predominantly to the excitation of magnon states, whereas the Sk state leads to the excitation of Sk type states, i.e. excitation of Sk and FM states concern very different states in the excited state spectrum of the system and thus lead to very different IETS spectra. Of particular interest is the excitation of the Sk state. It is very different for a central or peripheral injection of a tunneling electron, and this shows up in the junction conductance: for a central excitation, it leads to the extra-skyrmion states revealed by the present quantal study; the quantal nature of the system allows for spin singularities different from the 'usual' skyrmion state and these dominate the IETS spectrum for a central excitation. For a lateral excitation, different states are excited, corresponding to the variety of states with a skyrmionic singularity in the system

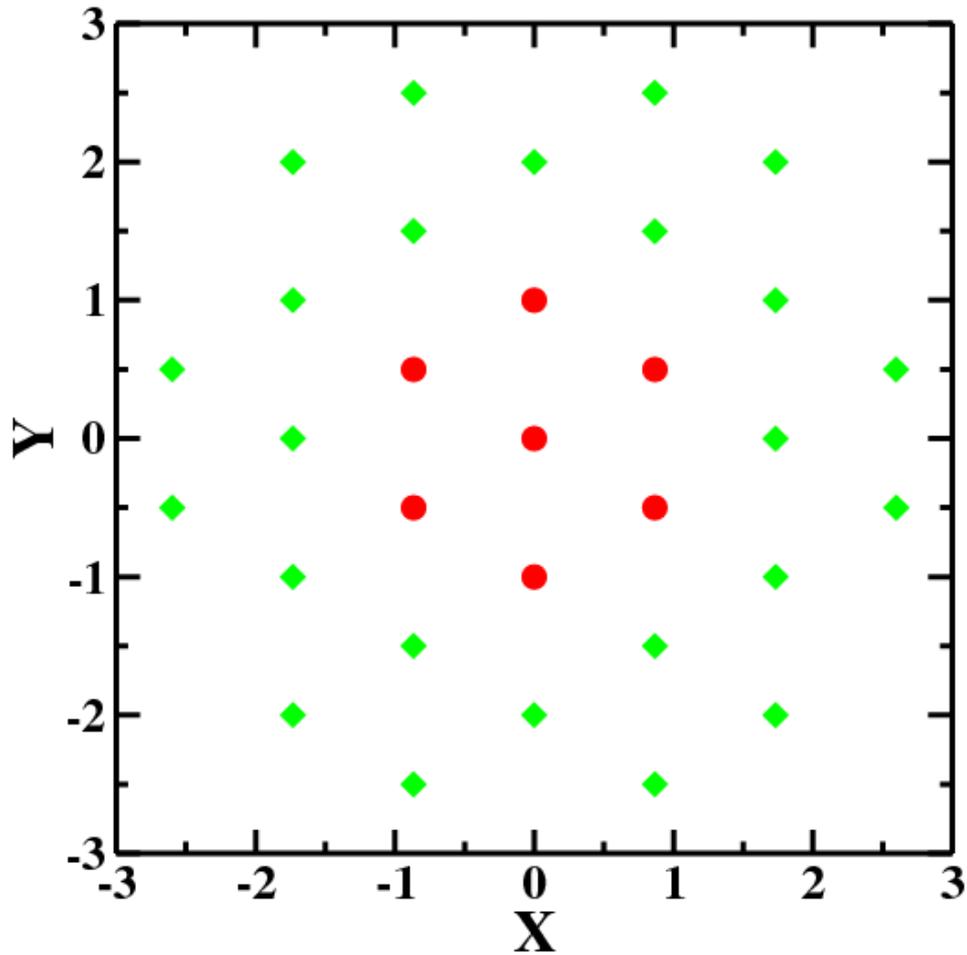

**Figure 1**: Schematic picture of the studied system. The seven central spins (red dots: central and first crown sites) are treated explicitly while the surrounding green dots sites provide a ferromagnetic embedment to the system. The position of the various sites are shown in the (X,Y) plane of the structure, with the unit length equal to the distance between two neighboring sites.

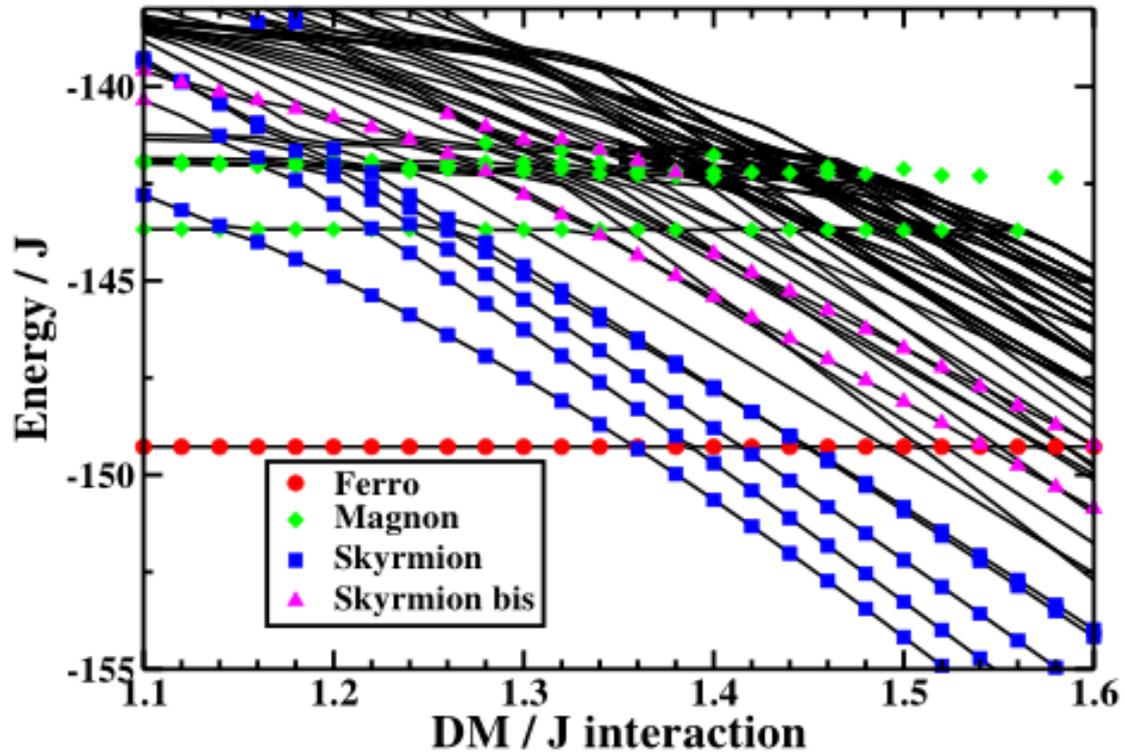

**Figure 2**: Energy of the 40 lowest states of the system as a function of the DM interaction (the 4-spins interaction is set equal to 0.1 J). Red dots: ferromagnetic state; Green diamonds: magnonic states; blue squares: skyrmion state (central spin up); purple triangles: skyrmion state with the central spin dominantly in the $S_z = +1/2$ state. The phase transition occurs around DM / J = 1.36 where the energies of the ferromagnetic and skyrmion states cross.

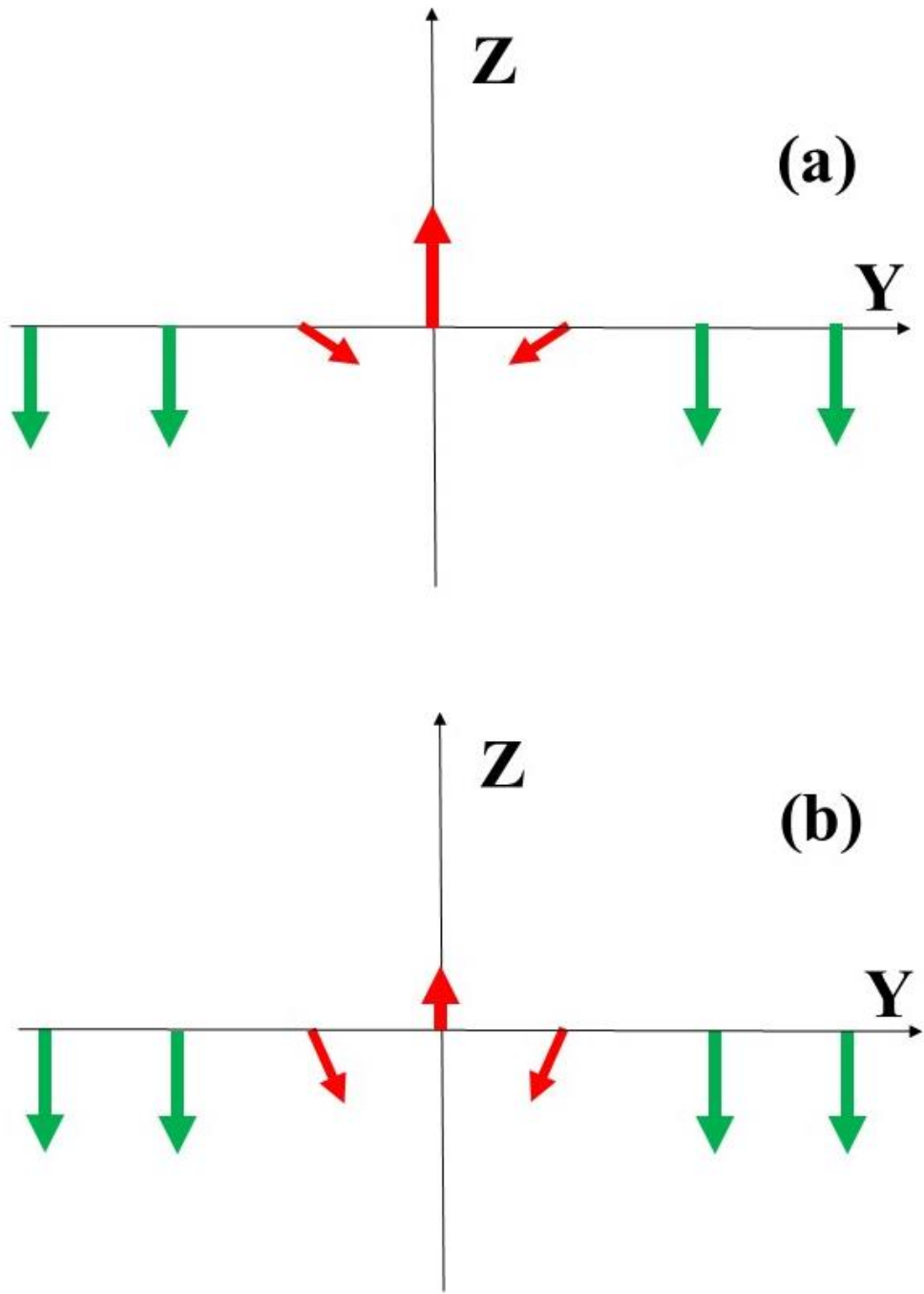

**Figure 3:** Schematic picture of a cut of the spin structure of the system in the (y,z) plane. The atoms are located on the y axis and their spins are in the (y,z) plane. The arrow represents the mean value of the spin operator at the site. The green arrows symbolize the spins of the ferromagnetic embedment (spin projection on the z-axis in the $m = -3/2$ state) and the red arrows the spins of the central structure. The DM interaction is set to 1.4 J and the 4-spins interaction to 0.1 J, i.e. the system is close to the FM-Sk phase transition. Part (a): structure of the ground skyrmion state. Part (b): structure of an excited skyrmion state (extra state with a spin projection on the z–axis dominantly in the $m = +1/2$ state on the central site).

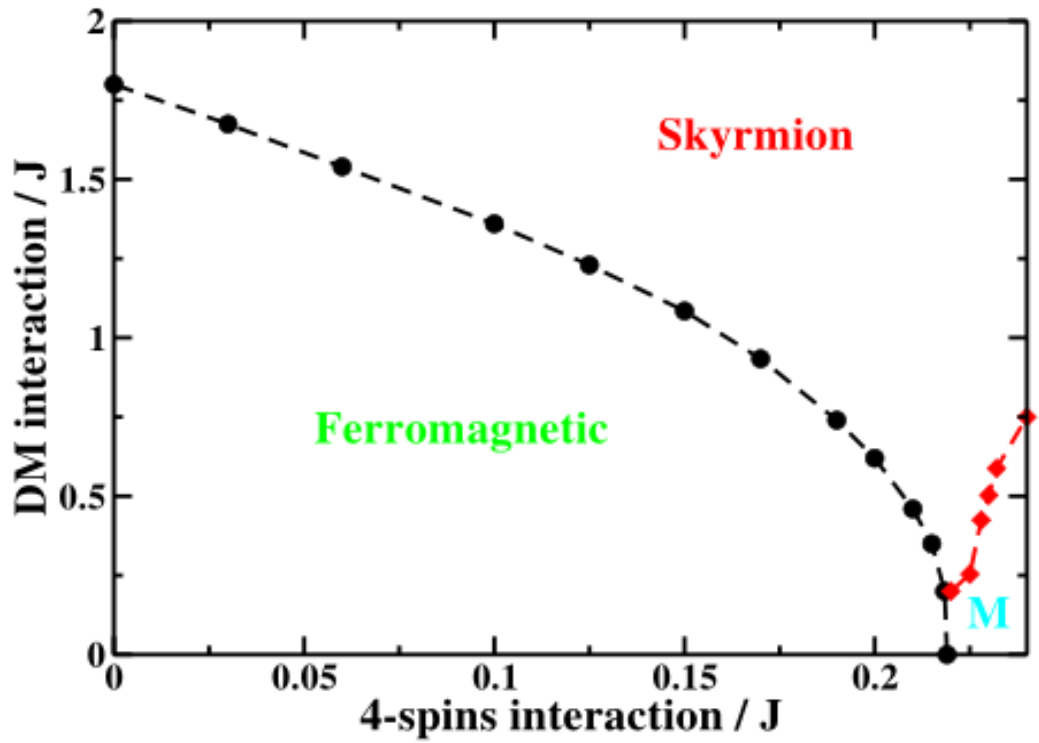

**Figure 4:** Phase diagram of the system in the (DM,4-spins) plane. Below the black dashed line the ground state is the FM state. Below the red-dashed line, the system is in a magnon-type state and above the two curves in a skyrmion state.

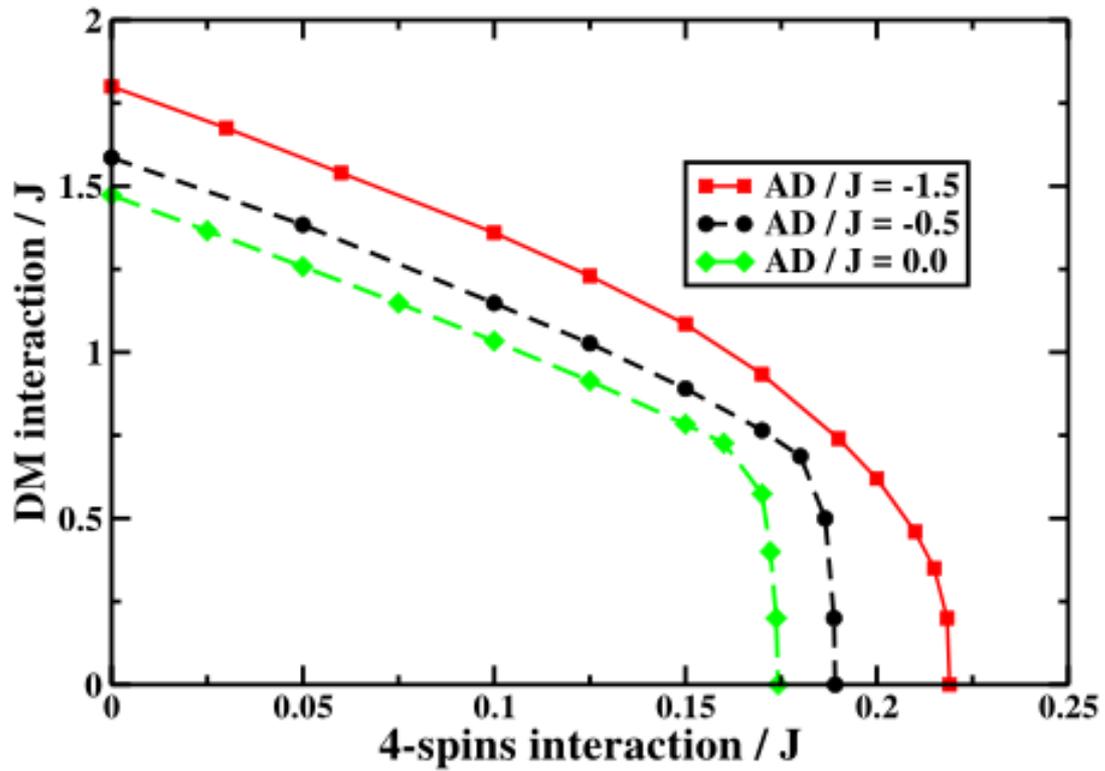

**Figure 5:** Phase diagram of the system in the (DM,4-spins) plane with different values of the on-site magnetic anisotropy AD. Only the FM-phase disappearance line is shown: the lowest energy state is the ferromagnetic state for parameters below the transition line. The three lines correspond to different values of the magnetic anisotropy.

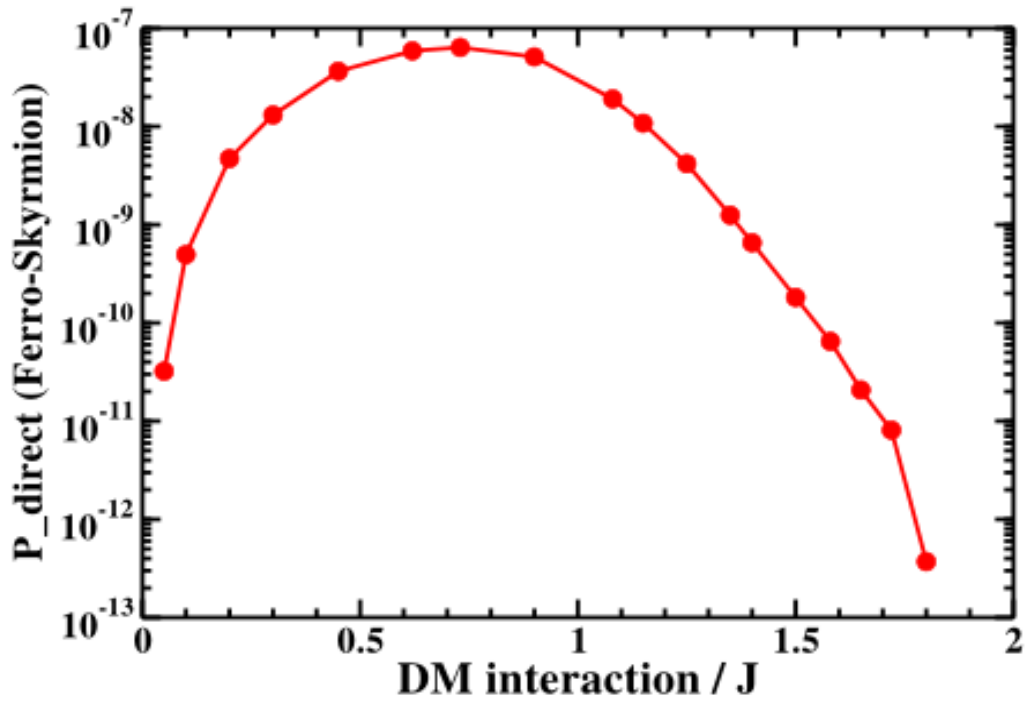

**Figure 6**: Transition probability for a direct transition between the ferromagnetic state and the lowest skyrmion state, $P_{direct}(FM \leftrightarrow Sk)$, induced by collision of an electron on one of the crown sites, as a function of the DM interaction. For each point, the 4-spin interaction is chosen so that the system is close to the phase transition, so that the figure actually presents the variation of $P_{direct}(FM \leftrightarrow Sk)$ along the FM-Sk transition line of Figure 4.

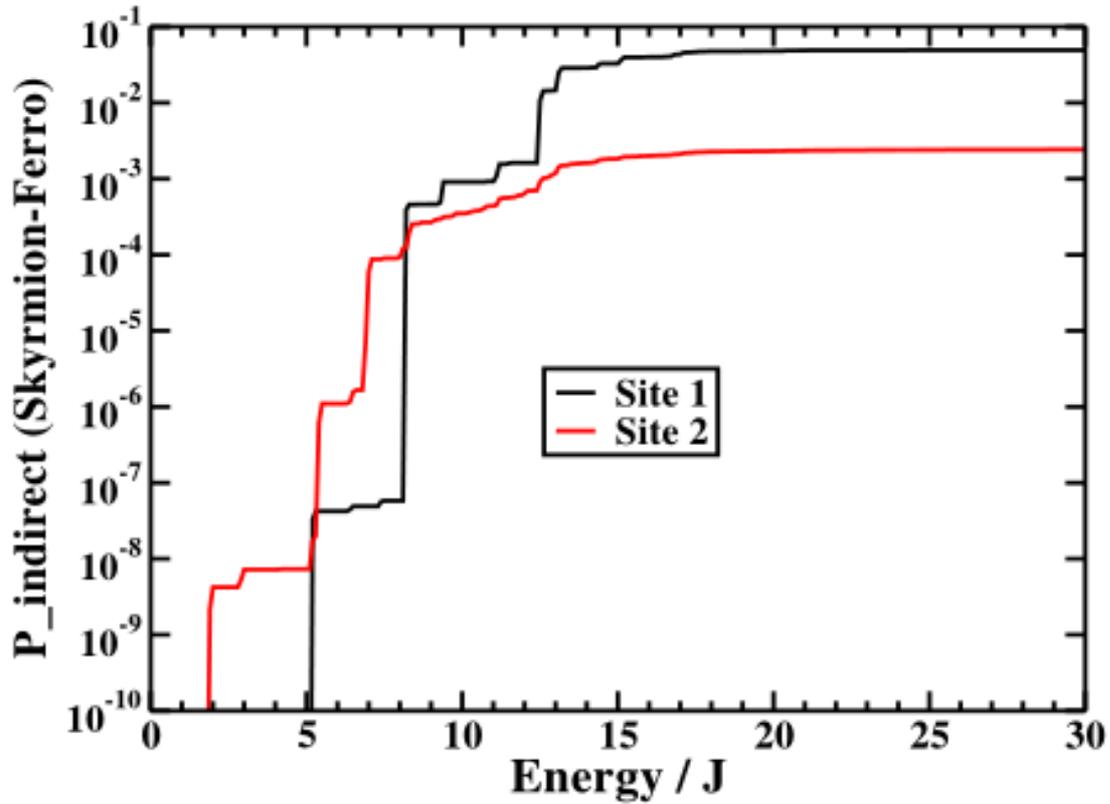

**Figure 7**: Transition probability for an indirect transition from the lowest skyrmion state to the ferromagnetic state, $P_{Indirect}(Sk \rightarrow FM)$, induced by collision of an electron on one of the system spins. The probability is plotted as a function of the electron incident energy. The electron excites the system initially in the ground skyrmion state to an intermediate excited state that later decays toward the ferromagnetic state. The DM interaction is set to 1.4 J and the 4-spins interaction to 0.1 J. Site 1 is the central site and site 2 is one of the crown sites.

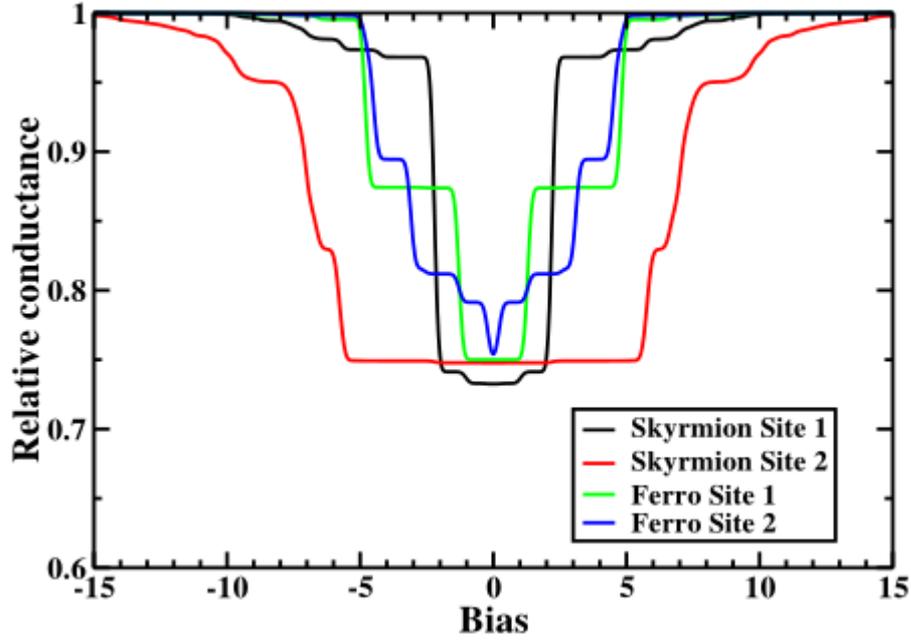

**Figure 8**: Relative conductance of the system for an electron injected into one of the atoms of the system (Site 1: central atom and Site 2: atom in the first crown). The conductance is normalized to one for large bias and the bias is expressed in units of e.mV/J. The system is initially in the ferromagnetic or lowest skyrmion state. The 4-spins interaction is set to 0.215 J and the DM interaction to 0.3 J. The ground state is ferromagnetic and the first excited state is a skyrmionic state i.e. the system is close to the phase transition with a small DM interaction.

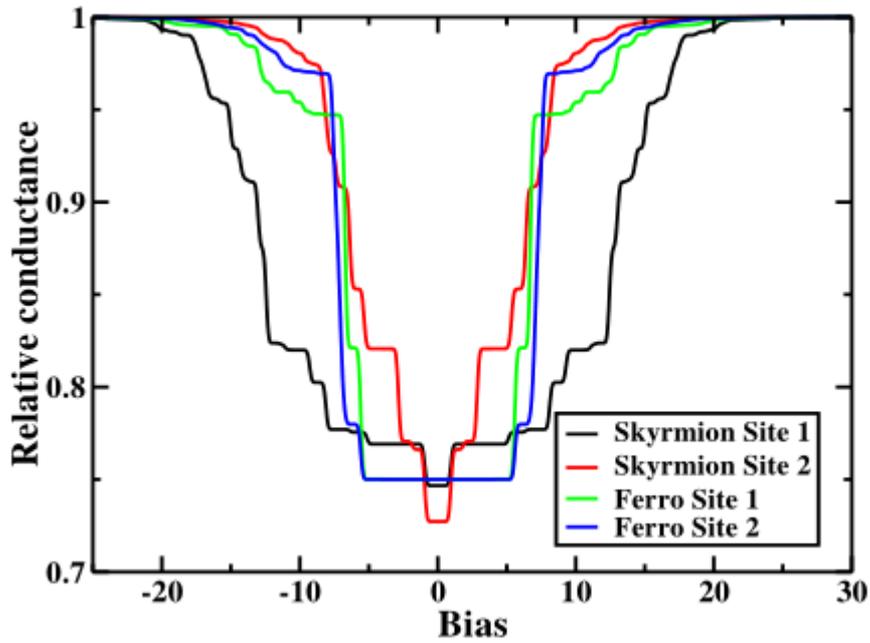

**Figure 9:** Relative conductance of the system for an electron injected into one of the atoms of the system (Site 1: central atom and Site 2: atom in the first crown). The conductance is normalized to one for large bias and the bias is expressed in units of e.mV/J. The system is initially in the ferromagnetic or lowest skyrmion state. The 4-spins interaction is set to 0.1 J and the DM interaction to 1.4 J. The ground state is a skyrmion state and the third excited state is the ferromagnetic state, i.e. the system is close to the phase transition with a large DM interaction (see Fig. 2).